\documentclass[aps,prb,twocolumn,longbibliography,superscriptaddress]{revtex4-1}

\usepackage{amsmath,amssymb}
\usepackage{graphicx}
\usepackage{epsfig, color}
\usepackage{wasysym}
\usepackage{amsfonts}
\usepackage{bm}
\usepackage{enumerate}
\usepackage{color}
\usepackage[resetlabels]{multibib}
\usepackage{epstopdf}
\usepackage{latexsym}
\usepackage[breaklinks,colorlinks = true,linkcolor = red,urlcolor=cyan,citecolor=red]{hyperref}
\usepackage[caption=false]{subfig}
\usepackage{float}

\newcommand{\bea}{\begin{eqnarray}}
\newcommand{\eea}{\end{eqnarray}}
\newcommand{\be}{\begin{eqnarray}}
\newcommand{\ee}{\end{eqnarray}}
\newcommand{\bw}{\begin{widetext}}
\newcommand{\ew}{\end{widetext}}

\usepackage[normalem]{ulem}

\def\ket#1{{|#1\rangle}}

\newcommand{\vs}[1]{\boldsymbol{#1}}
\newcommand{\abs}[1]{\left| #1 \right|} 

\begin{document}

\title{Topology and symmetry-protected domain wall conduction in quantum Hall nematics} 

\author{Kartiek~Agarwal}
\affiliation{Department of Physics, McGill University, Montr\'{e}al, Qu\'{e}bec H3A 2T8, Canada}
\affiliation{Department of Electrical Engineering, Princeton University, Princeton, New Jersey 08540, USA}
\author{Mallika~T.~Randeria}
\affiliation{Department of Physics, Princeton University, Princeton, New Jersey 08540, USA}
\author{A.~Yazdani}
\affiliation{Department of Physics, Princeton University, Princeton, New Jersey 08540, USA}

\author{S.~L.~Sondhi}
\affiliation{Department of Physics, Princeton University, Princeton, New Jersey 08540, USA}

\author{S.~A.~Parameswaran}
\affiliation{The Rudolf Peierls Centre for Theoretical Physics,  University of Oxford,  Oxford OX1 3PU, UK}

\date{\today}

\begin{abstract}
We consider domain walls in nematic quantum Hall  ferromagnets predicted to form in multivalley semiconductors, recently probed by scanning tunnelling microscopy experiments on Bi(111) surfaces. 
We show that the domain wall properties depend sensitively on the filling factor $\nu$ of the underlying (integer) quantum Hall states.  For $\nu=1$ and in the absence of impurity scattering we argue that  the wall hosts a single-channel Luttinger liquid whose gaplessness is a consequence of valley and charge conservation. For $\nu=2$, it supports a two-channel Luttinger liquid, which for sufficiently strong interactions enters a  symmetry-preserving thermal metal phase with a charge gap coexisting with gapless neutral  intervalley modes. The domain wall  physics in this state is identical to that of a bosonic topological insulator protected by $U(1)\times U(1)$ symmetry, and we provide a formal mapping between these problems. We discuss other unusual properties and experimental signatures of these `anomalous' one-dimensional systems.
\end{abstract}
\maketitle

\section{Introduction}

Topology and symmetry play central and intertwined roles in  condensed matter physics. In  Landau theory, different ordered phases are associated to distinct broken symmetries, with magnetism being the canonical example. Topology is then used to classify defects --- such as vortices, disclinations, or dislocations --- whose proliferation destroys order and restores symmetry. On the other hand the modern theory of topological states of matter distinguishes zero-temperature phases by global properties of their quantum wave functions, even in the absence of any symmetries --- as  most famously exemplified by two-dimensional electron gases (2DEGs) exhibiting the quantum Hall (QH) effect. When such  phases also spontaneously break symmetry, the interplay of broken symmetry and topological order can lead to new routes to stabilizing and manipulating topological phenomena.
 

Quantum Hall ferromagnets (QHFMs) furnish one such example, where the  formation of a topological QH state is driven by interaction-induced spontaneous breaking of a global symmetry, such as that associated with electron spin, or valley or layer pseudospin~\cite{SondhiSkyrmion}. QHFMs thus exhibit manifestations of both topological order---notably, quantized response and a vanishing energy gap for edge transport---as well as classic broken-symmetry phenomena, e.g. Goldstone modes and finite-temperature phase transitions ~\cite{Moon:1995p1}. Topological defects  gain additional structure from  the topological order of the underlying QH state---e.g., in spin QHFMs, skyrmion textures bind quantized electrical charge and can dominate low-energy charge properties~\cite{SondhiSkyrmion}. Studying these unusual topological defects can yield insight into an array of phenomena emerging from the interplay of interactions, symmetry, and topology.

Here, we focus on a particularly rich class of QHFMs, where the symmetry in question permutes distinct minima (`valleys')  of the low-energy electronic dispersion~\cite{Shkolnikov:2005p1, Mansour, IQHE_2007,mcfarland_temperature-dependent_2009,Padmanabhan:2010p1,Gokmen:2010p1,kott_valley-degenerate_2014}. Such systems~\cite{SAPBEFReview} are best described~\cite{abanin_nematic_2010} as {\it discrete nematics}: QH states with a symmetry-breaking order parameter that breaks the discrete rotational symmetry of the crystalline point group, and whose natural topological defects are domain walls, introduced e.g. by spatially varying uniaxial strain~\cite{Gokmen:2010p1}. 
Such a nematic QH liquid was recently observed via high-field scanning tunneling microscopy (STM) experiments on the sixfold valley-degenerate (111) surface of  bismuth (Bi)~\cite{Feldman316Full}. Orientational symmetry breaking is detected by imaging local density of states (LDOS) modulations near  atomic-scale impurities, while energy-resolved measurements clarify the role of interactions. 
Similar studies have now been performed at isolated domain walls between distinct nematic regions in the interior of a sample, far from physical edges~\cite{randeria2019interacting}. These reveal gapless domain wall modes  when the bulk QH  state is at Landau level filling factor $\nu=1$ but a tunneling gap when it is at $\nu=2$.

Usually, metallic conduction along edges of QH systems is protected by the fact that chiral edge modes transport charge unidirectionally; in contrast, at domain walls, one-dimensional (1D) charge modes counter-propagate. Since position and momenta are locked in the QH regime, interactions can strongly couple such counter-propagating modes without any constraints from momentum conservation, and it is natural to expect these modes to become gapped and insulating. 
New ideas are therefore necessary to explain the dichotomy between the tunneling spectra at different filling factors. 
Accordingly, we  develop a theory of  electronic degrees of freedom at these domain walls. We find that like in many 1D systems the relevant theory is that of a (multi-component) Luttinger liquid, but one in which  interactions are  constrained by momentum conservation in {\it two} dimensions (that origin of the valley symmetry). This structure is peculiar to the QH setting: such symmetry constraints cannot emerge in a local 1D quantum system. We develop these ideas quantitatively, place them within the framework of quantum anomalies, and use them to both explain STM data and explore their further implications.



\section{Microscopic Model}

\begin{figure}
\includegraphics[width=\columnwidth]{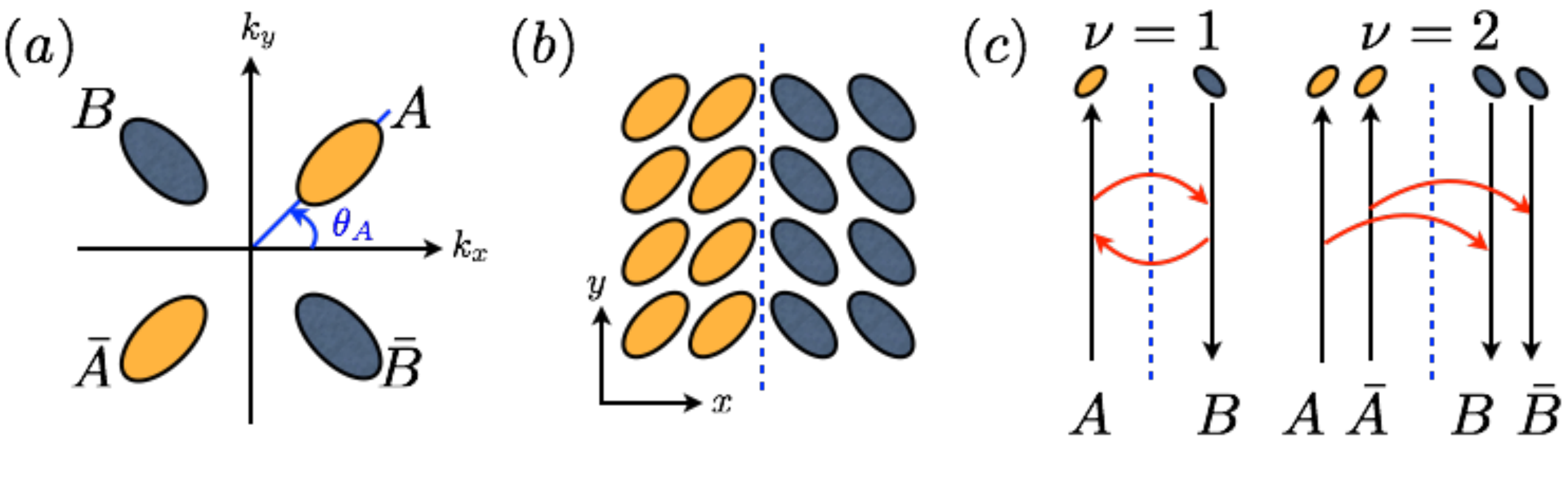}
\caption{\label{fig:bandsanddw} (a) Four-valley model. (b) Sketch of nematic domain wall. (c) Symmetry-allowed interactions at $\nu=1$ map to  forward scattering and cannot open a gap; example of allowed process that can open a charge gap at  $\nu=2$.}
\end{figure}

We study  a 4-valley model 
(Fig.~\ref{fig:bandsanddw}) of spin-polarized electrons described in a continuum effective mass approximation (valid when $\lambda_F, \ell_B\gg a $, where $\lambda_F$ is the Fermi wavelength, $a$ is of the order of the lattice spacing, and $\ell_B = \sqrt{\hbar/eB}$ is the magnetic length); we discuss later how to adapt this to Bi(111), which has 6 valleys. 
 The mass tensor is generically anisotropic, but respects $C_4$ point-group symmetry, so that discrete spatial rotations also permute valley indices. 
 We consider integer filling factors 
$ \nu_T =4p + \nu$, 
where $p$ is a nonnegative integer and $\nu=1,2$. ($\nu=3$ maps to $\nu=1$ under the `particle-hole' transformation.) For simplicity, we will also restrict to the lowest Landau level (LLL; $p=0$) though our results can be generalized to any $p$.  (barring competing density-wave instabilities which may be relevant for $p\geq 3$) 

The single-particle Hamiltonian for valley $\alpha \in \{A, B, \bar{A}, \bar{B} \}$ can be approximated as 
\begin{align}
H_\alpha = \frac{(p_\parallel - K + e A_\parallel / c)^2}{2 m_\parallel} + \frac{(p_\perp + e A_\perp / c)^2}{2 m_\perp},
\end{align}
 where $v_\parallel = v_x \cos \theta_\alpha + v_y \sin \theta_\alpha$, $v_\perp = v_y \cos \theta_\alpha - v_x \sin \theta_\alpha$ for any vector $\vs{v}$, and 
 $\theta_\alpha$ are angles shown in Fig.~\ref{fig:bandsanddw}. The valleys are centered at
 $\vs{K}_\alpha = K (\cos \theta_\alpha, \sin \theta_\alpha)$
and we define $\vs{K}_{\alpha \beta} = \vs{K}_\alpha - \vs{K}_\beta$.
 We assume that deviations from ellipticity (e.g., from the teardrop shape of Bi(111) valleys) denoted $\delta H_\alpha$,  are smaller than the mass anistropy $\lambda^2 = m_\parallel/m_\perp$; we discuss their role further in Appendix~\ref{app:corrections}.
 Working in Landau gauge $\vs{A} = (0, Bx)$, and introducing a guiding center $X$ related  to the momentum via  $X = \ell^2_B p_y$, yields single-particle wavefunctions $\phi_{\alpha, X} (\vs{r})$  in valley $\alpha$ 
 \begin{equation}\label{eq:LLLorb}
\phi_{\alpha, X} (x,y) = \frac{e^{i X y + i \vs{K}_\alpha \cdot \vs{r} }}{\sqrt{L_y}} \left( \frac{z'_\alpha}{\pi} \right)^{1/4} e^{-\frac{z_\alpha (x + X)^2}{2}}, 
\end{equation}
where $L_y$ is the length of the QH sample in the $y$-direction, $\lambda^2 = m_\parallel/m_\perp$ is the mass anisotropy, $z_\alpha = \frac{\lambda}{\lambda^2 \sin^2 \theta_\alpha + \cos^2 \theta_\alpha} + i \frac{\sin 2 \theta_\alpha ( 1- \lambda^2)}{2 (\lambda^2 \sin^2\theta_\alpha + \cos^2 \theta_\alpha)}$, and $z'_\alpha = \text{Re} \left[ z_\alpha \right]$.

 

Each non-interacting LL has an exact four-fold valley degeneracy. Therefore the formation of incompressible QH states for integer $\nu<4$ requires interactions; projecting these into the LLL yield the effective Hamiltonian
 \begin{equation} \label{eq:Hi}
H_i = \frac{1}{2A}\!\!\sum_{\vs{q}\alpha\beta\gamma\delta X X'} \!\!V(q) \; \vs{:}\!\bar{\rho}_{\alpha \beta} (\bar{\vs{q}}_{\alpha \beta}, X) \bar{\rho}_{\gamma \delta} (-\bar{\vs{q}}_{\delta \gamma}, X')\!\vs{:}.\!\!
\end{equation}

Here, $ \vs{:} \ldots  \vs{:} $ denotes normal ordering, $V (q)$ is the Fourier transform of the interaction. In terms of creation operators $c^\dagger_{\kappa, X}$ which create an electron in the LLL orbital $\phi_{\kappa, X}$, the density at wave-vector $q$, projected into the LLL is given by $\bar{\rho} (\vs{q}) = \sum_{\alpha \beta X} F_{\alpha \beta} (\vs{q}, X) \bar{\rho}_{\alpha \beta} (\vs{q}_{\alpha \beta}, X)$, where 
\begin{align}
\bar{\vs{q}}_{\alpha \beta} &= \vs{q} + \vs{K}_{\alpha \beta},  \\ 
\bar{\rho}_{\alpha \beta} (\bar{\vs{q}}_{\alpha \beta} , X) &= F_{\alpha \beta} (\bar{\vs{q}}_{\alpha \beta} , X) c^\dagger_{\kappa, X - \frac{\bar{q}_{y, \alpha \beta}}{2}}  c_{\kappa', X + \frac{\bar{q}_{y, \alpha \beta}}{2}}, \nonumber \\
F_{\alpha \beta} (\vs{q}, X) &= e^{i q_x X} \frac{(4 z'_\alpha z'_{\beta})^{1/4}}{\sqrt{z^*_\alpha + z_{\beta}}} e^{-  \frac{( q_x + i z^*_\alpha q_y) (q_x - i z_{\beta} q_y)}{2 ( z^*_\alpha + z_{\beta})} }\nonumber.
\label{eq:FF}
\end{align}

 \subsection{Hierarchy of terms}

The `form factors' $F_{\alpha\beta}(\vs{q})$ are exponentially sensitive to the momentum difference between the valleys $\alpha, \beta$. Accordingly, at leading order we may restrict to  
\begin{align}
H_{i,0} \boldsymbol{:} \; \text{terms in} \; H_i, \; \text{for} \; \alpha = \beta \; \; \gamma = \delta. 
\end{align}

Going to higher order, we find that  valley mixing interactions corresponding to near zero {total} momentum transfer in the 2D Brillouin zone 
 are only polynomially suppressed in $a/\ell_B $. Such terms fall into two categories:
 \begin{align}
 H_{i,1} \boldsymbol{:} \; \text{terms in} \; H_i, \; \text{for} \; (\gamma \delta) = (\beta \alpha), \nonumber \\
 H_{i,2} \boldsymbol{:} \; \text{terms in} \; H_i, \; \text{for} \;  (\gamma \delta) = (\bar{\alpha} \bar{\beta}). 
 \end{align}
 
Note that for both of the above terms, $\bar{q}_{\delta \gamma} = \bar{q}_{\alpha \beta}$. Then, a transformation $\vs{q} \rightarrow \vs{q} + \vs{K}_{\beta \alpha}$ transfers all dependence on $K$ into the argument $V(q)$, leading to an overall factor of $O(a/\ell_B)$ relative to $H_{i,0}$. In both $H_{i, 1}, H_{i,2}$ we require $\beta \neq \alpha$, and additionally in $H_{i,2}$, $\beta \neq \bar{\alpha}$. 

All other terms describe scattering processes with a large net 2D momentum transfer. While these are allowed in principle because of LLL projection, they are exponentially small $\sim e^{- (K \ell_B)^2} \approx e^{- (\ell_B/a)^2}$ and can be neglected. Thus, valley symmetries emerge as good approximate symmetries (see below). Strain will generically split the valley degeneracy fully at single particle level, but at leading order valleys $A, \bar{A}$ are approximately degenerate.  
 Note also that a strain field will generically split the valley degeneracy fully at single particle level, but at leading order valleys $A, \bar{A}$ are approximately degenerate and split only by $\delta H_\alpha$, as are $B, \bar{B}$;  we term these `anisotropy pairs'. 
For notational convenience, we dub the degree of freedom between two valleys that share the same anisotropy for $\delta H_{\alpha}=0$  (i.e., $X\leftrightarrow \bar{X}$ for $X=A,B$) `pseudospin' and that between such anisotropy pairs ($A\leftrightarrow B$), `isospin'. Domain walls between QHFMs polarized in different valleys are pinned by strain, that we model as a slowly varying valley Zeeman field that couples only  to isospin.
%
 
 \subsection{Symmetries}
In the elliptical-valley limit,  $\delta H_{\alpha}=0$, $H_{i,0}$ is invariant under $SU(2)$ pseudospin rotations. This yields a rich symmetry structure~\cite{QHFMOBD} but for our discussion we take $\delta H_\alpha \neq0$ (as is likely case in Bi(111)). However we will approximate the form factors by (\ref{eq:FF}). 
$H_{i,0}, H_{i,1}$ enjoy an emergent $[U(1)]^4$ symmetry, namely independent conservation of the electron number $N_\alpha$ in each valley.  (we assume $\delta H_{\alpha}$ also respects this).
 We can rearrange these into the following $4$ $U(1)$ charges
 \begin{align}
 \mathcal{N} &= N_A + N_B + N_{\bar{A}} + N_{\bar{B}},  \\
 \mathcal{P}^z &= \frac{1}{2}(N_A + N_B - N_{\bar{A}}- N_{\bar{B}}), \\
 \mathcal{I}^z &= \frac{1}{2}(N_A- N_B + N_{\bar{A}} - N_{\bar{B}}),  \\
 \mathcal{Q}^z  &= \frac{1}{2}(N_A - N_B - N_{\bar{A}} + N_{\bar{B}}). 
 \end{align}
These correspond to the total charge, $\mathcal{N}$, generators of rotations about the $z$-axes in pseudospin space, $\mathcal{P}^z$, and isospin space $\mathcal{I}^z$, and simultaneously in both, $\mathcal{Q}^z$. $H_{i,2}$ preserves $\mathcal{N}$, $\mathcal{P}^z$, and $\mathcal{Q}^z$, but breaks isospin $U(1)$ to $Z_2$, by allowing $ A\bar{A} \leftrightarrow B\bar{B}$ processes that change $\mathcal{I}^z$ in units of two. We will use these symmetries below to strongly constrain terms allowed in the low-energy theory of the domain wall.
 We comment here that the model has enhanced symmetry in the elliptical valley limit  $\delta H_{\alpha}=0$, where $H_{i,0}$ is invariant under $SU(2)$ pseudospin rotations. The rich symmetry structure~\cite{QHFMOBD} in this case may lead to additional interesting effects; however here we assume that $\delta H_\alpha \neq0$

 \subsection{QHFM ground states at $\nu = 1,2$.}
 
Ignoring intervalley contributions from $H_{i,1,2}$, at $\nu=1$ a Hartree-Fock (HF) calculation indicates electrons are polarized entirely in one of the valleys, $\ket{\Psi} = \prod_X c^\dagger_{\alpha, X}  \ket{0}$. Inter-valley coherent states that mix isospins are suppressed by the `large' anisotropy~\cite{abanin_nematic_2010} present already in the elliptical approximation, while pseudospin-mixing states are suppressed by the smaller anisotropy captured by $\delta H_{\alpha}$,  in accord with the microscopic symmetry~\footnote{Determining the ground state can be more subtle for $\delta H_{\alpha} =0$, since here the symmetries force an exact degeneracy between the $\nu=2$ with both valleys in an anisotropy pair filled, and one which corresponds to a $\nu=1$ state within each anisotropy pair. Selection between such states depends either on additional quantum fluctuations from $\delta H_\alpha$, thermal fluctuations of Goldstone modes from the broken $SU(2)$ symmetry of states within an anisotropy pair (`order by disorder') or by energetics of the flanking Wigner-crystal phases in clean systems (`order by doping')~\cite{QHFMOBD}. However, these subtleties are largely avoided by considering nonzero $\delta H_\alpha$ and also due to the presence of a symmetry-breaking strain field.}. Bulk excitations far from the wall are gapped for $\delta H_{\alpha} \neq 0$.
%
%
The relevant topological defects in this system, and our focus below, are isospin domain walls where the QHFM order parameter switches between anisotropy pairs. These can be induced by a spatially-varying uniaxial strain that splits isospin states at the single-particle level (but couples negligibly to pseudospin); this is captured by a parameter $\Gamma$ in our model, which we take to characterize the strain graident near the domain wall centre (where the strain vanishes). 

For $\nu=2$, we focus on pseudospin-singlet states where both partners in an anisotropy pair are occupied; a strain field will lower the energy of one anisotropy pair relative to the other so that an isospin domain wall again forms where the strain changes sign. 

Absent interactions, our model has 4 $U(1)$ symmetries, associated with charge conservation in each of the 4 valleys independently. At domain walls, these charges, which are associated with anomalous (quantum Hall) response in the bulk, give rise to gapless chiral edge modes by the Callan-Harvey mechanism~\cite{CALLAN1985427}. 
Interactions break some of these symmetries, and depending on 
the filling factor $\nu$ (defined modulo $4$ which corresponds to the filling of all $4$ valleys in a given Landau level), the residual symmetries suffice to protect some or all of the gapless domain wall modes. In what follows, we construct a Luttinger liquid theory for these domain walls using only symmetry arguments. A more microscopic calculation of the parameters is discussed in the Appendix~\ref{app:micro}. 



\section{Domain Walls at $\nu = 1$: symmetry-protected metallic state} 
\subsection{Luttinger Liquid Description}
At $\nu = 1$, only valleys $A, B$ are occupied and $H_{i,2}$ is thus irrelevant. 
We therefore only consider the remaining valley-$U(1)$-conserving  interactions, $H_{i, 0,1}$ 
and the smoothly varying valley Zeeman field $\Delta_v$ that energetically stabilizes the domains.

Without loss of generality, we will assume that on the left of the domain wall, valley $A$ is occupied (states $X < 0$), and on the right, valley $B$ is occupied (states $X > 0$).  
For  $\Delta_v=0$, the domain wall has a zero mode corresponding to a rigid translation of the wall~\cite{Mitra:2003p1,FernBondesanSimon}. 
Microscopically, this mode changes a fixed number of left moving electrons into right movers. This corresponds to the transformation $\rho_R \rightarrow \rho_R + \epsilon, \rho_L \rightarrow \rho_L - \epsilon$, where we identify $\rho_r (q_y) \sim \sum_X c^\dagger_{X + q_y, \alpha (r)} c_{X, \alpha (r)}$, with $r = L,R$ labeling left/right moving electrons, and corresponding valley indices $\alpha(L) = A, \alpha (R) = B$.

Since $H_{i,0}, H_{i,1}$ respect this symmetry corresponding to the free translation of the domain wall transverse to itself, the corresponding terms in the effective Hamiltonian must take the form 
\begin{align}
H_0 \equiv H(\Gamma = 0) = \pi v^0_F \int dy \; \left[ \rho_R (y) + \rho_L (y) \right]^2,
\label{eq:nu1H01}
\end{align}
where $v^0_F$ is a renormalized effective velocity and $\Gamma$ parametrizes the gradient in $\Delta_v$ (see Appendix~\ref{app:micro}). 
Note that the effective Hamiltonian corresponding to the valley Zeeman field is a single-particle term that corresponds exactly to the usual Tomonaga-Luttinger electron gas, and thus has the form 
\begin{align}
H_v = \pi \Gamma \int dy \; \left[ \rho^2_R (y) + \rho^2_L (y) \right],
\label{eq:nu1Hv}
\end{align}

Writing the densities in terms of the canonically conjugate fields $\phi, \pi \Pi$, with $\nabla \phi = - \pi \left[ \rho_L + \rho_R \right], \Pi = \rho_R - \rho_L$~\cite{Giamarchibook}, we find the effective Luttinger liquid Hamiltonian for the $\nu = 1$ domain wall
 \begin{eqnarray}
H_{\text{DW}}^{\nu=1} = \frac{u}{2 \pi} \int dy \; \left[ \frac{1}{K} (\nabla \phi)^2 + K (\pi \Pi)^2 \right].
\label{eq:HDW1}
\end{eqnarray}
 Here $K = \sqrt{\Gamma / (v^0_F + \Gamma)}$, $u = \sqrt{v^0_F \Gamma} \sqrt{1 + \Gamma/v^0_F}$ are strain-dependent and vanish for $\Gamma = 0$, reflecting the zero mode in the limit $\Gamma \rightarrow 0$. 

Note that unlike usual 1D systems such as nanotubes, here scattering between left- and right-moving states  involves \emph{no} change in momentum along the wall, since the position-momentum locking in the LL ensures that states at the same guiding center $X$ are proximate in momentum $p_y$. Naively, it seems that interactions could then lead to a quantum-disordered gapped phase as $T\rightarrow 0$. However, here the valley momentum difference $\vs{K}_{A B}$ ensures that such processes are in fact suppressed exponentially, hence the domain wall remains gapless. The chiral modes in each direction carry distinct valley quantum numbers; this valley-filtered nature provides an intuitive explanation for the symmetry protection. 

\subsection{Symmetry Analysis and Gapping Perturbations}
The Luttinger Hamiltonian in Eq.~(\ref{eq:HDW1}) describes a gapless system, where the Luttinger parameter $K$ grows smaller (and thus interactions grow stronger) as the valley Zeeman field grows weak (smaller $\Gamma$). We now discuss potential gapping perturbations and identify the symmetries that forbid these.  

We use the standard bosonization dictionary for spinless electrons, where $ \psi_{r}(y) = \frac{U_{r,s}}{\sqrt{2\pi\alpha}}e^{-\frac{i}{\sqrt{2}}[r\phi(y) - \theta_(y)]}$ where $r=\pm$ for  $R,L$, and we map $\{L, R\} \equiv \{A, B\}$. In this notation, we have
\begin{subequations}
\begin{eqnarray}
\nabla \phi &= -\pi [ \rho_{R} + \rho_{L}] = -\pi [ \rho_{B} + \rho_{A}],  \\
\nabla \theta &= \pi [ \rho_{R} - \rho_{L}] = \pi [ \rho_{B} - \rho_{A}]
\end{eqnarray}
\end{subequations}
Cosines and sines of linear combinations of these phases $\phi, \theta$ then comprise the usual gapping perturbations of the system. In terms of these, the conserved charges may be obtained by integrating appropriate linear combinations:
\begin{subequations}
\begin{eqnarray}
 \mathcal{N} &=& N_A +N_B= -\frac{1}{\pi}\int dy\, (\nabla \phi) \\
 \mathcal{I}^z &=& \frac{1}{2}(N_A - N_B) = -\frac{1}{2\pi}\int dy\, (\nabla \theta)
\end{eqnarray}
\end{subequations}
where we use the fact that $N_{\alpha} \sim \int dy \rho_\alpha$. Further, recall the commutation relation 
\begin{equation}
[ \phi(y),\frac{1}{\pi}\nabla \theta(y')] = [ \theta(y),\frac{1}{\pi}\nabla \phi(y')]=  i\delta(y-y'), 
\end{equation}
Using the identity that $[A, e^{iB}] = i [A,B]e^{iB}$ for $[[A,B],B]=0$, we find
\begin{subequations}
\begin{align}
[\mathcal{N}, e^{\pm i\theta(y)}] &= \mp e^{\pm i\theta(y)}, \\
[\mathcal{I}^z, e^{\pm i\phi(y)}] &= \mp e^{\pm i\phi(y)}. 
\end{align}
\end{subequations}
Thus, $e^{\pm i\theta}$ and $e^{\pm i\phi}$ correspond to lowering/raising operators for the charges $\mathcal{N}$, $\mathcal{I}^z$. Since these operators are conserved in our system, any operator built from these is forbidden by symmetry. As a consequence, there are no symmetry-allowed perturbations to the $\nu=1$ domain wall, which is thus always in a gapless phase as long as charge and valley $U(1)$ symmetries are preserved.

To see how the symmetry protection is linked to the topological response, imagine applying an electric field parallel to the domain wall. Owing to the QH response, this induces an electrical current perpendicular to the wall. Therefore, electrons in valley $A$   flow towards the wall from the left, and valley $B$ electrons  flow away from it on the right. As long as the valley quantum number is conserved, there is then a net current of valley {isospin} {\it into} the wall. If the wall were insulating, this would lead to an inconsistency: it must therefore carry gapless {isospin} excitations. Reversing this argument, we see that a `{isospin} field' parallel to the wall (i.e., a positive (negative) electric field for valley $A$ ($B$)) would drive charge current into the wall.  Thus,  the domain wall excitations are also electrically charged. This can be viewed as the  Callan-Harvey `anomaly inflow' argument~\cite{CALLAN1985427} adapted to the QHFM setting. 


\subsection{Alternative approach via nonlinear sigma model}
For completeness, we present an alternative discussion of protected conduction at $\nu=1$ in a field-theoretic framework that is often used to discuss  quantum Hall ferromagnets. For the nematic case, this takes the form of an easy-axis non-linear sigma model for the ferromagnetic order parameter, $\vec{m}  = (m_x,m_y,m_z)$ (with $\vec{m}^2=1$):
\begin{align}
\mathcal{S} [\vec{m}] &= \mathcal{S} _B[\vec{m}] +\mathcal{S} _{\text{g}}[\vec{m}]+ \mathcal{S}_{\text{QH}}[\vec{m}],
\end{align}
where the first term
\begin{equation}
 \label{eq:Berry}
\mathcal{S} _B =  \int d^2r d\tau\,  i S\vec{\mathcal{A}}[\vec{m}]\cdot\partial_\tau{\vec{m}}
\end{equation}
 is the standard Berry phase kinetic term for a ferromagnet, with $\nabla_{\vec{m}}\times\mathcal{A}[\vec{m}] = \vec{m}$ and $S=1/2$,  and
 \be 
\mathcal{S} _{\text{g}}[\vec{m}] =\int d^2r d\tau\, \left[ \frac{\rho_s}{2} \left(\nabla m\right )^2 + \frac{\alpha}{2}\vec{m}_\perp^2 +  \Gamma x  m_z(\vec{r})\right]\,\,\,
\ee
is the usual gradient energy of an easy-axis sigma model with stiffness  and easy-axis anisotropy $\alpha>0$, and we have included a ``Zeeman gradient'' $\Gamma$; for the moment we ignore possible anisotropic stiffness terms as they only give small corrections. The gradient term this has $[U(1)\rtimes \mathbb{Z}_2]_s$ pseudospin rotation symmetry, where the $U(1)$ corresponds to rotations about the $m_z$-axis, and the $\mathbb{Z}_2$ takes $m_z\rightarrow -m_z$, and the semidirect product $(\rtimes)$ indicates that these two operations do not commute. 

In addition are two special features of this nonlinear sigma model due to the underlying quantum Hall physics, captured in $\mathcal{S}_{\text{QH}}$. First, textures of the  order parameter $\vec{m}$ with a nonzero Pontryagin index  $C= \int \frac{d^2 r}{4\pi}\, \epsilon_{abc}\epsilon^{\mu\nu}{m^a}(\nabla_\mu m^b)(\nabla_\nu m^c)$   carry an electric charge $Q = \nu eC$ (To avoid confusion with the $U(1)_s$ spin rotation,  we will refer to the corresponding charge conservation symmetry as $U(1)_c$); second, $\mathcal{S}_{\text{QH}}$ also contains topological `Hopf' term that does not have a simple local expression~\footnote{Although it does have a simpler expression in terms of the alternative $CP^1$ formulation of the nonlinear sigma model.} in terms of $\vec{m}$. This is essentially the transcription of the Chern-Simons term of the underlying  quantum Hall state and at $\nu=1$ endows the skyrmions with fermionic statistics. The explicit form of  $\mathcal{S}_{\text{QH}}[\vec{m}]$ is not particularly important, but its physical manifestations  --- namely that skyrmions are fermions with unit $U(1)_c$ charge --- are crucial in distinguishing the QHFM from a conventional ferromagnet, and will be significant when discussing gapping perturbations to the domain wall.

 
However, as a first step, let us  ignore $\mathcal{S}_{\text{QH}}$; deriving the effective domain wall theory is then a standard exercise in  soliton dynamics via the method of classical coordinates. A slowly fluctuating domain-wall solution takes the form $\vec{m} = (\sin\theta\cos\phi, \sin\theta\sin\phi, \cos\theta)$ with
\be
\phi(\vec{r},t) = \phi(y,t), \,\,\, \cos\theta(\vec{r},t) = \tanh\frac{x-X(y,t)}{\lambda}.
\ee
Here, the classical `soft' coordinates are the phase along the wall $\phi(y,t)$  (note that this is only  meaningful near the wall) and the location of the center of the wall $X(y,t)$ (defined implicitly by requiring  $m_z(X(y,t)) = 0$). Taking $\phi = X = \text{const.}$ corresponds to a static saddle-point whose energy is minimized for $\lambda = \sqrt{\rho_s/\alpha}$, and (for any $\Gamma\neq 0$) $X=0$. Note that the saddle-point energy is independent of $\phi$ and so it seems that our choice spontaneously  breaks the $U(1)_s$ symmetry; in higher dimensions there would be a Goldstone mode associated with this, but of course this is precluded in 1+1D by the Mermin-Wagner theorem, and fluctuations restore symmetry. The low-energy effective dynamics at the domain wall are captured by a 1+1D action for $\phi, X$ that may be obtained by performing a gradient expansion  in fluctuations of the `slow fields' $\phi$, $X$ about the static saddle point. 
%
This yields the  domain wall effective action 
\be
\mathcal{S} = \int\!dyd\tau \left[  i{2S} \dot{\phi} X+  \Gamma X^2 + \beta (\nabla X)^2 + \frac{\rho}{2} (\partial_y\phi)^2 +\ldots \right]\nonumber\\
\ee
where $\beta, \rho >0$ are constants whose precise value is unimportant and $\ldots$ indicates higher-order terms. 
Note that in the absence of a pinning term $\Gamma$, we have a $z=2$ theory (this can be  verified by computing the equations of motion for $\beta\neq 0$):  exactly as we found for the microscopic model for $\Gamma\rightarrow 0$. Since we do have pinning, we may set $\beta=0$ and integrate out the fluctuations of the domain wall position $X$. This yields the phase-only effective action
\be
\mathcal{S}_\phi = \int dy d\tau\,\left[ \frac{S^2}{\Gamma} (\partial_\tau \phi)^2 + \frac{\rho_s}{2} (\partial_y \phi)^2 \right]
\ee
yielding a Luttinger liquid where both $u, K \propto \Gamma^{1/2}$ consistent with the  more microscopic approach.

At this level, it seems that it should be possible to disorder this 1D theory by breaking the $U(1)_s$ symmetry, either fully by adding, e.g. a term $\propto m_y$, or down to $Z_2$, by adding a term $\propto m_y^2$, respectively resulting in $\delta\mathcal{S}_\phi \propto \cos \phi, \cos 2\phi$. If we demand that $U(1)_s$ is preserved, we may rule out such terrms. However, in a 1+1D quantum theory, we  can also gap the system by a quantum vortex unbinding transition, corresponding to driving the dual field $\cos\theta$ to strong coupling. For an ordinary  ferromagnet (that is, if we ignore $\mathcal{S}_{\text{QH}}$) nothing seems to obstruct this transition: there is no reason {in principle} to forbid a trivial, gapped (quantum-disordered) phase of the domain wall modes.
 
\begin{figure}[t!]
\includegraphics[width=\columnwidth]{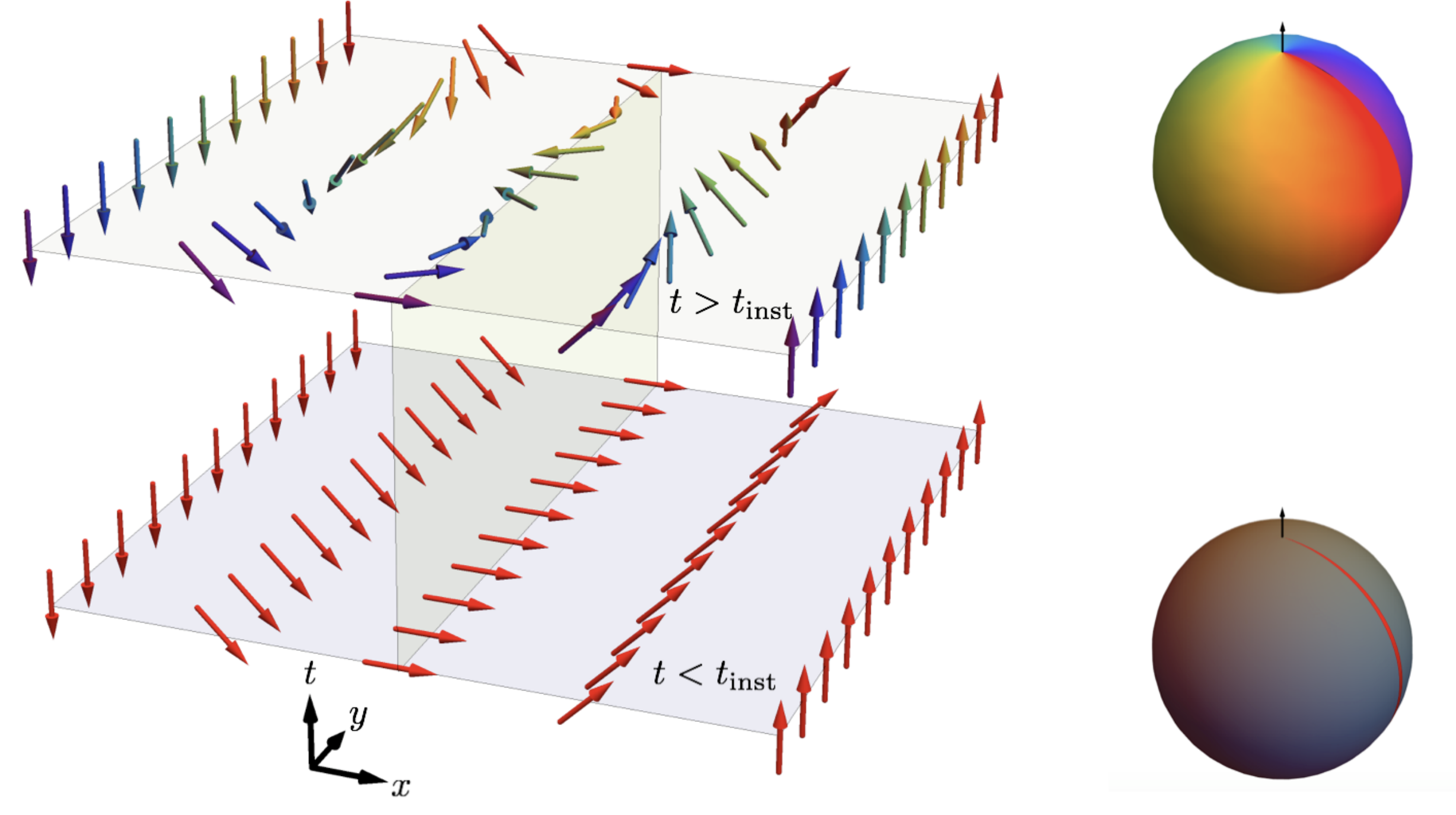}
\caption{\label{fig:skvort} A spacetime vortex in the phase on the domain wall worldsheet (yellow) can be viewed as a $2\pi$ kink inserted instantaneously at $t= t_{\text{inst}}$. Since spins in the bulk on either side of the DW are oppositely oriented, for $t<t_{\text{inst}}$ when there is no kink at the DW, the spin configuration in the 2D plane is defined by a great semi-circle on the Bloch sphere that passes through the equator at $\phi=0$ (bottom right). For $t>t_{\text{inst}}$, owing to the presence of the kink,  the spin configurations wrap the sphere as we move parallel to the domain wall (the color of the arrows indicates the azimuthal angle on the Bloch sphere, top right). The skyrmion number of the 2D spin configuration thus increases by 1 each time a vortex event (instanton) occurs.}
\end{figure}

 This conclusion is altered by including the effects of $\mathcal{S}_{\text{QH}}$. A space-time vortex is an instanton: a quantum process that inserts  a $2\pi$ kink in the phase winding at some instant in time. However, we must remember that the domain wall does not exist in isolation: it is flanked by two different orientations of the $Z_2$ part of the pseudospin. So, if we view the 2D system {\it after} the insertion of the $2\pi$ kink at the domain wall,  there is now a skyrmion present in the system. Therefore, we see that the spacetime vortex operator in the 1+1D theory is a skyrmion creation operator (Fig.~\ref{fig:skvort}) (Note: this is also consistent with the view that in spacetime this defect is a hedgehog, since spacetime hedgehogs change skyrmion number by 1.) Because of the topological terms discussed above, this carries both a unit $U(1)_c$ charge, and fermion parity. Therefore, we see that the 1+1D QHFM domain wall theory is unconventional: its instanton defects carry electrical charge and fermion number. Single defect proliferation is therefore forbidden, as it would violate the fermion parity symmetry. Double-defect proliferation is consistent with fermion parity, but would break $U(1)_c$ down to $\mathbb{Z}_2$, corresponding to superconducting pairing. Therefore, we conclude that the gapless Luttinger liquid at the domain wall is symmetry-protected: while the $U(1)_s$ protection is present in any NLSM (even for a conventional ferromagnet), the $U(1)_c$ protection is unique to the QHFM, and is important in ruling out a trivial gapped phase. In this fashion, the robust gapless edge mode protection is ultimately a consequence of underlying QH anomaly, and hence can be argued to be the same as the Callan-Harvey mechanism. It is clearly also consistent with the microscopic symmetry analysis above.

\section{Domain walls at $\nu = 2$: charge-insulating thermal metal} 
\subsection{Luttinger Liquid Description}
We may proceed analogously for the $\nu = 2$ case and construct an effective Luttinger liquid description using symmetry arguments. As mentioned above, we assume that a valley Zeeman field, which does not distinguish between pseudospin valley pairs (such as $A$, $\bar{A}$), varies spatially, and stabilizes the occupation of states $A, \bar{A}$ on the left of the system (states $X < 0$), and $B, \bar{B}$ on the right (states $X> 0$). In this case, there are a pair of counterpropagating edge modes, one from each of the two filled Landau levels. Noting the valley index of the left- and right-movers, we write 
\begin{equation}\{ A, \bar{A}, B, \bar{B} \}, \equiv \{ (L, \uparrow), (L, \downarrow), (R, \uparrow), (R, \downarrow) \},\end{equation} tracking the valley polarization on either side of the wall. 

Next, using standard arguments~\cite{Giamarchibook} we  decouple `charge' and `valley' sectors whose densities are given by sum and difference of opposite pseudospin ($\uparrow / \downarrow$) densities. Since $H_v, H_{i,0}, H_{i,1}$ preserve the $4$ valley $U(1)$ symmetries, these terms do not comprise gapping perturbations. Thus, the effective Hamiltonian corresponding to these terms is given by a sum of two Luttinger Hamiltonians, that is, $H = H_\rho +H_\sigma$ with
\begin{eqnarray}
H_{\zeta} &=& \frac{ u_{\zeta} }{2 \pi} \int d y \; \left[ \frac{1}{K_{\zeta}} \left( \nabla \phi_{\zeta} \right)^2 + K_{\zeta} \left( \pi \Pi_{ \zeta} \right)^2  \right],
\label{eq:HchargespinLuttinger}
\end{eqnarray}
where the charge and valley modes are denoted $\zeta = \rho, \sigma$, and have distinct Luttinger parameters in general. Crucially, by similar arguments as above, we note that these parameters much depend singularly on $\Gamma$ which parameterizes the valley Zeeman field. (The lack of a stabilizing valley Zeeman field must yield a zero mode corresponding to translations, and in this limit the Luttinger parameter is zero.) 

Now, at $\nu=2$, valley-mixing interactions play a crucial role: $H_{i,2}$ (that describes a scattering process involving electrons in all four valleys) leads to a backscattering interaction in the charge sector given by 
\begin{align}
H_{i,2} = \frac{2}{(2 \pi \alpha)^2} \int dy \; \! \text{Re}[ g e^{i \sqrt{8} \phi_\rho} ]. 
\label{eq:Hi2}
\end{align}
(The precise form of this term is guaranteed by the action of the operator $e^{\pm i \sqrt{2} \phi_\rho}$ which changes the charge $\mathcal{I}_z$ by units of $2$, which is the defining feature of $H_{i,2}$.) Now, combining all the terms, the effective Hamiltonian is  $H_{\text{DW}}^{\nu=2} = H_\rho + H_\sigma + H_{i,2}$. 

The relevance of the backscattering interaction depends on the value of the charge Luttinger parameter $K_\rho$. For repulsive interactions and weak strain gradient $\Gamma$, generically we find 
$K_\rho \ll 1$ so that $H_{i,2}$ is always relevant~\cite{Giamarchibook}. Thus the theory is driven to strong coupling, pinning  $\phi_\rho$ to a minimum of the cosine. This disorders $\theta_\rho$, i.e. the correlation function $\langle e^{\frac{i}{\sqrt{2}}\theta_\rho({x,t})}e^{\frac{i}{\sqrt{2}}\theta_\rho({0,0})}\rangle$ decays exponentially. Since $e^{-\frac{i}{\sqrt{2}}\theta_{\rho}}$ is related to charge creation, we see that now charge correlations decay along the wall, which is thus electrically insulating. In contrast, the excitations in the $\sigma$ channel remain gapless. The domain wall is thus fractionalized in the sense that the charge is frozen while the valley degrees of freedom propagate freely. We note further that when $H_{i,2}$ is relevant, the ground state of the cosine potential has minima $\phi_\rho = \bar{\phi}_\rho + \frac{2 n \pi}{\sqrt{8}}$, for $n \in \mathcal{Z}$ of which one is chosen; in Appendix~\ref{app:compact} we show these minima correspond to the same physical state. 

\subsection{Symmetry Analysis and Gapping Perturbations}
We now perform a symmetry analysis analogous to the case $\nu =1$. Following our conventions  and the mapping $\{ A, \bar{A}, B, \bar{B} \} \equiv \{ (L, \uparrow), (L, \downarrow), (R, \uparrow), (R, \downarrow) \}$ we may write
\begin{subequations}
\begin{align}
\nabla \phi_{\uparrow} &= -\pi [ \rho_{R,\uparrow} + \rho_{L,\uparrow}] =-\pi [ \rho_{B} + \rho_{A}],  \\
\nabla \theta_{\uparrow} &= \pi [ \rho_{R,\uparrow} - \rho_{L,\uparrow}] =\pi [ \rho_{B} - \rho_{A}],  \\
\nabla \phi_{\downarrow} &= -\pi [ \rho_{R,\downarrow} + \rho_{L,\downarrow}] =-\pi [ \rho_{\bar{B}} + \rho_{\bar{A}}],  \\
\nabla \theta_{\downarrow} &= \pi [ \rho_{R,\downarrow} - \rho_{L,\downarrow}] =\pi [ \rho_{\bar{B}} - \rho_{\bar{A}}].
\end{align}
\end{subequations}
In terms of these, the conserved charges may be obtained  by integrating appropriate linear combinations:
\begin{subequations}
\begin{eqnarray}
 \mathcal{N} & = -\frac{1}{\pi}\int dy\, (\nabla \phi_\uparrow + \nabla\phi_\downarrow)  = -\frac{\sqrt{2}}{\pi}\int dy\, \nabla \phi_\rho\,\,\,\,
 \\
 \mathcal{P}^z &= -\frac{1}{2\pi}\int dy\, (\nabla \phi_\uparrow - \nabla\phi_\downarrow)= -\frac{1}{\sqrt{2}\pi}\int dy\, \nabla \phi_\sigma\,\,\,\,
 \\
 \mathcal{I}^z &=  -\frac{1}{2\pi}\int dy\, (\nabla \theta_\uparrow + \nabla\theta_\downarrow) =  -\frac{1}{\sqrt{2}\pi}\int dy\, \nabla \theta_\rho\,\,\,\,
 \\
 \mathcal{Q}^z  &=  -\frac{1}{2\pi}\int dy\, (\nabla \theta_\uparrow - \nabla\theta_\downarrow) =-\frac{1}{\sqrt{2}\pi}\int dy\, \nabla \theta_\sigma\,\,\,\,
\end{eqnarray}
\end{subequations}

where we again use the fact that $N_{\alpha} \sim \int dy \rho_\alpha$. The commutation relation  is now
\begin{equation}
[ \phi_\eta(y),\frac{1}{\pi}\nabla \theta_{\eta'}(y')] = [ \theta_\eta(y),\frac{1}{\pi}\nabla \phi_{\eta'}(y')]=  i\delta_{\eta\eta'}\delta(y-y'), 
\end{equation}
where $\eta,\eta' \in \{\rho,\sigma\}$. Proceeding as for the $\nu=1$ case, we find
\begin{subequations}
\begin{align}
[\mathcal{N}, e^{\pm \frac{i}{\sqrt{2}}\theta_\rho(y)}] &= \mp  e^{\pm \frac{i}{\sqrt{2}}\theta_\rho(y)}\\
[\mathcal{P}^z, e^{\pm i\sqrt{2}\theta_\sigma(y)}] &= \mp e^{\pm i\sqrt{2}\theta_\sigma(y)} \\
[\mathcal{I}^z, e^{\pm i\sqrt{2}\phi_\rho(y)}] &= \mp e^{\pm i\sqrt{2}\phi_\rho(y)}\\
[\mathcal{Q}^z, e^{\pm i\sqrt{2}\phi_\sigma(y)}] &= \mp e^{\pm i\sqrt{2}\phi_\sigma(y)}
\end{align}
\end{subequations}
with all other commutators with conserved charges being zero. Thus we see that $e^{\pm \frac{i}{\sqrt{2}}\theta_\rho(y)}$, $e^{\pm i\sqrt{2}\theta_\sigma(y)}$, $e^{\pm i\sqrt{2}\phi_\rho(y)}$, $e^{\pm i\sqrt{2}\phi_\sigma(y)}$ (note the factors of $\sqrt{2}$) are respectively lowering/raising operators for $\mathcal{N}, \mathcal{P}^z, \mathcal{I}^z, \mathcal{Q}^z$. [The operator $e^{\pm i\sqrt{2}\theta_\rho(y)}$ changes $\mathcal{N}$ by two units; this is consistent since such an operator is produced by `pairing' bilinears of the form $\psi^\dagger\psi^\dagger$, while any single-electron annihilation has the form $\psi \propto e^{\frac{i}{\sqrt{2}}\theta_\rho}$.]

Since  $\mathcal{N}, \mathcal{P}^z, \mathcal{Q}^z$ are good quantum numbers, all cosines of the form $\cos(n\sqrt{2}\theta_\rho)$, $\cos(n\sqrt{2}\theta_\sigma)$, and $\cos(n\sqrt{2}\phi_\sigma)$ are forbidden for any $n$ as the corresponding operators break these symmetries. However, processes that change $\mathcal{I}^z$ in units of two are allowed, by terms in $H_{i,2}$, and correspond to the $n=2$ operator $\cos(\sqrt{8} \phi_\rho)$. 


The above analysis further implies that the gaplessness of the valley mode is robust and protected by this triplet of $U(1)$ symmetries.  A topological argument for the presence of such a gapless mode may also be made for the $\nu = 2$ case. {Let us first consider the case where $H_{i,2}$ is not present, and each valley is associated with a conserved charge. The QH response of the bulk pairs up the charges $\mathcal{N}$ with $\mathcal{I}_z$, and $\mathcal{P}_z$ with $\mathcal{Q}_z$. To see this, note that an application of a fictitious field parallel to the domain wall, that couples directly to one of these charges, drives an accumulation of the complementary charge at the domain wall due to the bulk QH response. The latter necessitates the presence of a gapless mode along the domain wall to carry away the excess charge. As noted above, this is a straightforward generalization of the Callan-Harvey argument for the presence of gapless edge modes associated with conserved charges that exhibit topological response in the bulk; the subtlety here is that a field gradient of one charge appears to drive the accumulation of a \emph{complementary} charge at the boundary. Next, if we allow for $H_{i,2}$, there exists a process to convert a pair of charges from the valleys $A, \bar{A}$ to charges of valleys $B, \bar{B}$, and vice-versa. Now, the application of an electric field along the domain wall---which drives charges from valleys $A, \bar{A}$ ($ B, \bar{B}$) into (away from) the domain wall---does not lead to an accumulation of isospin charge at the domain wall because of the process mentioned above. The `charge' mode is thus not protected (and it is gapped). }

Such a situation, where all perturbations are forbidden based solely on symmetry without tuning parameters, is impossible in truly 1D systems. This, like the linking of valley index to chirality, is tied to the fact that QHFM domain walls are `anomalous' and  can only be realized in conjunction with a topologically ordered bulk~
, similarly to helical edge states in 2D quantum spin Hall insulators. 

\subsection{Link between $\nu=2$ domain walls and bosonic topological insulator}
It is useful also to briefly make a link~\footnote{We thank Yi-Zhuang You for drawing our attention to this point.} to a superficially very different problem: that of  bosonic symmetry-protected topological phases  protected by $U(1)\times U(1)$ 
symmetry. As argued in Ref.~\onlinecite{BLGBSPT}, absent interactions, undoped graphene bilayers can be driven into an analog of a quantum spin Hall (QSH)  state with two effective  helical modes at each edge. Each of these helical edge modes has up spins and down spins propagating in opposite directions. 
Since the magnetic field explicitly breaks time reversal symmetry, unlike the usual QSH insulator, this state is actually protected by a pair of $U(1)$ symmetries: total charge and total spin. Interactions can gap out the charge modes  thereby opening an electron spectral gap at the boundary. However, this leaves a protected  neutral bosonic mode --- whose symmetry protection follows because the possible cosines are ruled out respectively by the charge and spin $U(1)$ symmetries. The formal similarity between this problem and our domain wall system may be made more concrete by `folding' the system across the domain wall, i.e., by viewing the domain instead as a edge between a  quantum valley Hall state (where the $A, \bar{A}$ valleys see positive magnetic field and the $B, \bar{B}$ valley see a negative magnetic field) and the vacuum. After we write down the single gapping cosine, the remaining valley mode is protected precisely by a pair of $U(1)$ symmetries --- in our case, valley pseudospin and valley isospin.  We refer the reader to Ref.~\onlinecite{BLGBSPT} for a discussion of why it is reasonable to use the term `bosonic topological insulator' despite the fact that the fundamental particles in the system are electrons.

\section{Relation to STM Experiments}

We may directly validate  our analysis against the STM data on Bi(111), where strain splits the six valleys into a (4,2) degeneracy pattern. Our model captures the remaining 4-fold degeneracy, with mirror reflections  constraining dispersions rather than $C_4$. In the gapless $\nu=1$ case, ideal STM experiments will see a soft gap due to Luttinger liquid suppression, with an energy/temperature dependence set by the Luttinger parameter $K$~\cite{Giamarchibook}. However, it is likely challenging to resolve this in realistic experimental settings. For $\nu=2$ we expect a {\it hard} gap~\cite{VoitTunneling, EsslerTsvelikSpectralFunctions} owing to charge-valley separation, as can be seen by expressing the single-electron spectral function using $\theta_\eta, \phi_\eta$, and using the exponential decay of charge correlations~\cite{Giamarchibook}. 
Taking $\lambda=5$, and approximating screening crudely via a large dielectric constant $\epsilon \approx 45$, yields a bulk exchange gap~\cite{Feldman316Full} $\Delta_{\text{ex}} \sim 535\,\mu$eV, and Luttinger liquid parameters $u_\rho \sim 0.1 \Delta_{\text{ex}} \ell_B$, $K_\rho \sim 0.1$ for $\Gamma\sim 0.01 \Delta_{\text{ex}} \ell_B$. For $\nu=2$ we estimate a charge gap of $120\,\mu$eV for small $\Gamma$, a sizable fraction of ~$\Delta_{\text{ex}}$; this is is consistent with our discussion above and the dichotomy between $\nu=1,2$ reported in Ref.~\onlinecite{randeria2019interacting}.
\begin{figure}
\includegraphics[width=\columnwidth]{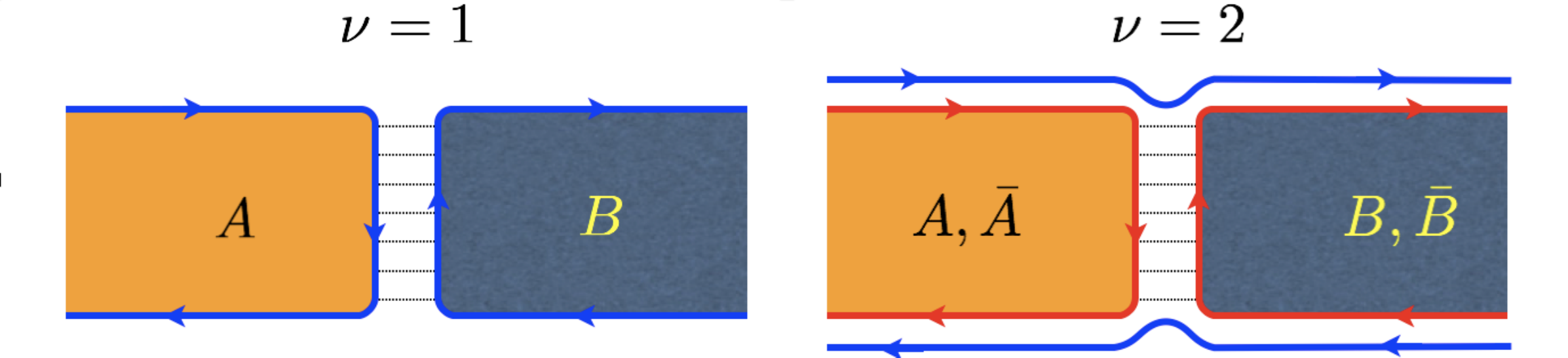}
\caption{\label{fig:linejunc} Domain walls as `line junctions'. For $\nu=1$ the wall conducts charge (blue) even with interactions (dashed lines); so, conductance is not quantized. For $\nu=2$, charge is gapped while neutral valley modes (red) are gapless, so that electrical (thermal) conductance is quantized (non-quantized).}
\end{figure}

\section{Concluding Remarks}
Motivated by our success in explaining STM data, we  now explore other implications of our theory. The key physical insight is that the $\nu=1$ domain wall is both electrically and thermally conducting, whereas the $\nu=2$ wall is a charge insulator but a thermal metal. This has immediate consequences for two-terminal measurements in the `line junction' limit~\cite{LineJunction,linejunction2} (Fig.~\ref{fig:linejunc}) with a single domain wall transverse to the direction of current flow. Namely, we expect no quantized conductance in the $\nu=1$ case since the wall transports charge between the edge modes, whereas for $\nu=2$ we expect the quantized charge conductance but no quantized thermal conductance. 

This observation generalizes to the phase diagram of the system in the presence of long-wavelength disorder, that preserves the valley symmetry crucial for the domain wall structure to survive. Assuming that the random strain produced by disorder vanishes on average, we expect large samples to contain many domains of the two possible nematic orientations, separated by a percolating network of domain walls. For  $\nu=1$ the gapless charge transport along the domain walls will lead to bulk dissipation and hence absence of a quantized Hall plateau~\cite{abanin_nematic_2010,KumarSAPSLSDWs}. At $\nu=2$, the charge gap leaves the quantization of charge Hall conductivity intact, but the gapless valley modes transport heat in the bulk. This destroys the quantization of the thermal Hall conductivity leading to an unusual violation of the Weidemann-Franz law, manifested in off-diagonal components of the conductivities. 
 Remarkably, in this scenario the interplay of disorder with the QHFM domain wall physics allows a familiar 1D effect --- `fractionalization' of transport --- to drive a similar phenomenon in 2D.
Uniform strain breaks the statistical symmetry between the domains, leading to a net excess of one domain over the other. In this regime, the domain wall network is tuned away from percolation, and no longer shorts the edges; in this limit, the quantized (thermal) Hall conductivity is restored at $\nu=1$ ($\nu=2$). We also expect various interesting but sample-dependent mesoscopic effects at intermediate scales. Other possibilities, e.g. localization of the domain-wall network, could lead to richer phenomenology, exploration of which we defer to the future.

As our arguments have built primarily on symmetry and topology, we expect that they will apply generally to a range of multivalley systems, such as graphene multilayers, transition metal dichalcogenides, and semiconductor heterostructures, particularly AlAs and Si based heterostructures which host six-fold symmetric valleys akin to Bi(111). In the former~\cite{shayegan2006two,herzog2012optimization}, the pseudospin pairs of valleys are identified by reciprocal lattice vectors. Thus our analysis for $\nu = 1$ is most relevant as there is effectively no pseudospin degree of freedom. In the latter, a $(4,2)$ valley degeneracy structure has been observed experimentally~\cite{eng2007integer} and our analysis should apply directly at fields where spin splitting is substantial. There are many avenues for further study; among them we flag especially the possibility of exploring similar phenomena in quantum magnetism, that has traditionally shared fruitful interactions with QH physics~\cite{KomargodskiWall}. Another exciting possibility is to extend our analysis to the fractional QH regime. Here, different candidate QH states, e.g. at $\nu=2/3$, may be distinguished via their domain-wall properties, a traditionally challenging problem; meanwhile,  the ability to introduce various gapping perturbations may allow domain walls to serve as a platform for engineering topologically protected qubits.

\section*{Acknowledgements} 
We thank J.T.~Chalker, F.H.L.~Essler, L.~Glazman, A.~Nahum, S.H.~Simon,  K.~Shtengel, and Y.-Z. You for  insightful and stimulating discussions. KA acknowledges support from the U.K. foundation and DOE DE-SC0002140. SAP acknowledges support from NSF DMR-1455366 during the early stages of this project and from  European Research Council (ERC) under the European Union Horizon 2020 Research and Innovation Programme [Grant Agreement No~804213-TMCS] as this work was completed. MTR and AY acknowledge support from the Gordon and Betty Moore Foundation as part of EPiQS initiative (GBMF4530), DOE-BES grant DE-FG02-07ER46419, and NSF-MRSEC programs through the Princeton Center for Complex Materials DMR-142054, NSF-DMR-1608848.

\appendix

\section{\label{app:corrections}
Role of Corrections to Ellipticity}

In writing the microscopic wavefunctions, we assumed that the valleys are perfectly elliptical. In reality, there can be corrections beyond ellipticity -- e.g. valleys in Bi have a `teardrop' shape.  We assume that deviations from ellipticity (e.g., from the teardrop shape of Bi(111) valleys) denoted $\delta H_\alpha$,  are smaller than the mass anistropy $\lambda^2 = m_\parallel/m_\perp$. 
Indeed our discussion we have implicitly assumed  a small but nonzero $\delta H_\alpha$ (as for Bi(111)), although we continue to approximate the form factors by (\ref{eq:FF}). 
This has two main consequences. First,  on the elliptical-valley limit,  $\delta H_{\alpha}=0$, $H_{i,0}$ is invariant under $SU(2)$ pseudospin rotations. This yields a rich symmetry structure that would complicate the discussion, in particular making the determination of a ground state at $\nu=1,2$ much more subtle~\cite{QHFMOBD}. Formally the terms in $\delta H_\alpha$ lowers this SU(2) symmetry to $\mathbb{Z}_2$: therefore, they suppress pseudospin-coherent ground states at $\nu=1$, in favor of states where the pseudospin is maximally polarized into one or other member of an anisotropy pair. 
Furthermore, $\delta H_\alpha$ gaps bulk collective excitations far from the wall, and allows us to focus our attention on the domain wall.

\section{\label{app:compact}
Compactification at $\nu = 2$}

The cosine potential for the $\phi_\rho$ field has several minima---$\phi_\rho = \bar{\phi}_\rho + \frac{2 n \pi}{\sqrt{8}}$, for $n \in \mathbb{Z}$---of which one is chosen. Here we show that these minima correspond to the same physical state. 

As noted, absent interactions there are 4 $U(1)$ symmetries, each associated with the total charge in each valley: $N_A, N_B, N_{\bar A}, N_{\bar B}$. If these symmetries are not broken, then there must exist chiral fermionic modes at the edge of the sample, as guaranteed by the Callan-Harvey mechanism~\cite{CALLAN1985427}. These fermionic modes may be expressed in terms of chiral bosonic fields, that is, $\psi_A \sim e^{- i \varphi_A}, \psi_{\bar A} \sim e^{- i \varphi_{\bar A}}, \psi_B \sim e^{i \varphi_B}, \psi_{\bar B} \sim e^{i \varphi_{\bar B}}$, where these bosonic fields obey standard commutation relations~\cite{Giamarchibook}: $[\varphi_{r (\kappa)} (x), \varphi_{r (\kappa')} ] = i \pi r (\kappa) \delta_{\kappa \kappa'} \text{sgn} \left[ x - x' \right]$. Here $r(A) = r(\bar{A}) = -1$ and $r (B) = r(\bar {B}) = 1$. We may then rearrange these operators to arrive the field operators used in the main text: 

\begin{align}
\phi_\rho &= \frac{1}{\sqrt{2}} \left[ \frac{\varphi_A + \varphi_B}{2} + \frac{\varphi_{\bar A} + \varphi_{\bar B}}{2} \right], \nonumber \\
\phi_\sigma &= \frac{1}{\sqrt{2}} \left[ \frac{\varphi_A + \varphi_B}{2} - \frac{\varphi_{\bar A} + \varphi_{\bar B}}{2} \right], \nonumber \\
\theta_\rho &= \frac{1}{\sqrt{2}} \left[ \frac{\varphi_A - \varphi_B}{2} + \frac{\varphi_{\bar A} - \varphi_{\bar B}}{2} \right], \nonumber \\
\theta_\sigma &= \frac{1}{\sqrt{2}} \left[ \frac{\varphi_A - \varphi_B}{2} - \frac{\varphi_{\bar A} - \varphi_{\bar B}}{2} \right], 
\end{align}
and check that these satisfy the usual commutation relations noted above. One can further identify $\phi_\uparrow = - \left( \frac{\varphi_A + \varphi_B}{2} \right), \phi_\downarrow = - \left( \frac{\varphi_{\bar A} - \varphi_{\bar B}}{2} \right)$, $\theta_\uparrow = \left( \frac{- \varphi_A + \varphi_B}{2} \right), \theta_\downarrow =  \left( \frac{- \varphi_{\bar A} + \varphi_{\bar B}}{2} \right)$. Using the usual expression for the chiral density, $\rho_{\kappa} = \frac{1}{2\pi} \nabla{\varphi}_\kappa$, we can arrive at all the results of the previous section. 

We can now identify the compactification radius of $\phi_\rho$. Since $\varphi_\kappa$ are independent $U(1)$ phases with a compactification radius $2\pi$, that is, $\varphi_\kappa \equiv \varphi_\kappa + 2 \pi$, we note that $\phi_\rho$ must be identified with $\phi_\rho + \frac{2 n \pi}{\sqrt{8}}$ for $n \in \mathbb{Z}$. Thus, all the minima of the cosine potential correspond to the same physical state.

\section{\label{app:micro}Microsopic estimation of Luttinger parameters}

In the main text, we provided rigorous, symmetry-based arguments for our general expectations for domain wall excitations.  We now discuss a more microscopic procedure for constructing the Luttinger theory derived above. This will allow us to provide estimates of the Luttinger parameters relevant to experimental observations of such modes.   
\subsection{General philosophy}
We will focus on single-particle modes near the domain wall, that is, modes with $X, X' \approx 0$. For such modes, the effective interaction may be approximated as 
\begin{widetext}
\begin{align}
H_i &= \frac{1}{2A} \sum_{\alpha \beta \gamma \delta, X, X', \vs{q}} G_{\alpha \beta \gamma \delta} (q_y) \vs{:} c^\dagger_{\alpha, X - \frac{q_y}{2}} c_{\beta, X + \frac{q_y}{2}} c^\dagger_{\gamma, X' - \frac{q_y}{2}} c_{\delta, X' + \frac{q_y}{2}} \vs{:}, \nonumber \\
G_{\alpha \beta \gamma \delta} (q_y) &\equiv \int \frac{dq_x}{2 \pi} \; V_{\vs{q}} F_{\alpha \beta} (\bar{\vs{q}}_{\alpha \beta}, X = 0) F_{\gamma \delta} (- \bar{\vs{q}}_{\delta \gamma}, X' = 0) \approx G_{\alpha \beta \gamma \delta} (0) \,\,\forall\,\, q_y
\label{eq:Happrox}
\end{align}
\end{widetext}

We made two assumptions in the above result. First, we neglected $X,X'$ dependence in the effective interaction amplitude $G_{\alpha \beta \gamma \delta}$, confining our attention to the physics near $X = X' = 0$. Second, note that the interaction amplitude $G(q_y) \sim e^{-q^2_y/2}$. Thus, the effective interaction is Gaussian in momentum exchanged along the domain wall, and consequently it is also Gaussian in spatial extent along the domain wall. This suggests that it can be approximated by a contact interaction, which corresponds to setting $G(q_y) = G(0) \forall q_y$. Landau level projection thus naturally leads to a theory of fermionic modes propagating along the domain wall and interacting via short-range interactions.  

The density of left and right moving fermions in this system may be defined as discussed above: for $\nu = 1$, $\rho_L (y) = \frac{1}{L_y} \sum_{q_y} e^{i q_y y} \left[ \sum_X c^\dagger_{A, X} c_{A, X - q_y} \right]$, and $\rho_R (y) = \frac{1}{L_y} \sum_{q_y} e^{i q_y y} \left[ \sum_X c^\dagger_{B, X} c_{B, X - q_y} \right]$. One may similarly define densities for the $\nu = 2$ case. 

The above problem of fermions interacting with contact interactions can be treated analogous to Luttinger liquid analysis developed for spinless and spinful fermions, see Ref.~\onlinecite{Giamarchibook}. The various interaction amplitudes determine the effective Luttinger parameters of our theory. 

\subsection{$\nu = 1$.}
We assume, as above, that valley $A$ is occupied for $X < 0$, and valley $B$ is occupied for $X > 0$. A valley Zeeman field gradient, $\sum_X \Gamma X \left( c^\dagger_{A, X} c_{A, X} -  c^\dagger_{B, X} c_{B, X} \right)$, that supports such a domain wall configuration, then directly translates into the term $\int d y \; \pi \Gamma \left[ \rho_R^2 (y) + \rho_L^2 (y) \right]$. Thus, the parameter $\Gamma$ in $H_v$ [in Eq.~(\ref{eq:nu1Hv})] may be estimated directly by the gradient of the valley Zeeman field (which in turn can be estimated by the gradient of the strain field and the difference of its coupling to the different valley modes). Note that this follows analogously to a free Tomonaga-Luttinger gas noting that the guiding center $X$ is also the momentum of the orbital in the $y-$direction.  

For $\nu = 1$, $H_{i,2}$ is irrelevant. $H_{i,0}, H_{i,1}$ must respect free rigid translations of the domain wall, and therefore must be of the form given in Eq.~(\ref{eq:nu1H01}). The parameter $v^0_F$ can be read off by transforming appropriate terms in Eq.~(\ref{eq:Happrox}) into the Luttinger liquid variables. This yields
\begin{align}
v^0_F &= \frac{1}{2\pi} \int \frac{q_x}{2\pi} \; V (\vs{q}) \abs{F_{AA} (\vs{q})}^2 \large|_{q_y = 0} \nonumber \\
&- \frac{1}{2\pi} \; V \left(\vs{q} + \vs{K}_A - \vs{K}_B \right) \abs{F_{AB} (\vs{q})}^2 \large|_{q_y = 0}. 
\end{align}

\subsection{$\nu = 2$.}

The single-particle valley Zeeman term is given now by two copies of the Tomonaga-Luttinger Hamiltonian
\begin{align}
H_v &= \pi \Gamma \int dy \! \left[ (\rho^2_{L, \uparrow} + \rho^2_{R, \uparrow}) + (\rho^2_{L, \downarrow} +  \rho^2_{R, \downarrow}) \right].
\end{align}

As for $\nu = 1$, $\Gamma$ is directly given by the valley Zeeman field gradient. As before, the interaction terms allow for a rigid translation of the domain wall, and the corresponding terms in the effective Hamiltonian must reflect this symmetry. We further note that $H_{i,0}$ is symmetric with respect to all valleys, and thus leads to a term of the form 
\begin{align}
H_{i,0} &= \pi v^0_F \int dy \! \left[ \rho_{L} + \rho_{R} \right]^2, \nonumber \\
\rho_{L} &= \rho_{L, \uparrow} + \rho_{L, \downarrow}, \; \rho_{R} = \rho_{R, \uparrow} + \rho_{R, \downarrow}. 
\end{align}
where 
\begin{align}
v^0_F = \frac{1}{2\pi} \int \frac{dq_x}{2\pi} V (\vs{q}) \abs{F_{AA} (\vs{q}) }^2 \large|_{q_y = 0}. 
\end{align}

$H_{i,1}$ involves exchange interactions between valley pairs, with an amplitude that is generically different for pairs separated by a momentum shift along and/or against the domain wall. It transforms into
\begin{align}
H_{i,1} &= - \pi v^1_F ( 1 + \chi) \int dy \;\!\! \left[ (\rho_{L, \uparrow} + \rho_{R, \uparrow} )^2 + (\rho_{L, \downarrow} + \rho_{R, \downarrow})^2 \right] \nonumber \\
-& \pi v^1_F ( 1 - \chi)  \int dy \;\!\! \left[ (\rho_{L, \uparrow} + \rho_{R, \downarrow})^2 + (\rho_{L, \downarrow} + \rho_{R, \uparrow})^2 \right] \big\}
\end{align}
where 
\begin{align}
v^1_F (1 + \chi) = \frac{1}{2\pi} V \left(\vs{q} + \vs{K}_A - \vs{K}_B \right) \abs{F_{AB} (\vs{q}) }^2 \large|_{q_y = 0}, \nonumber \\
v^1_F (1 - \chi) = \frac{1}{2\pi} V \left(\vs{q} + \vs{K}_A + \vs{K}_B \right) \abs{F_{A \bar{B}} (\vs{q}) }^2 \large|_{q_y = 0}. 
\end{align}
The above expressions may then be converted into a usual Luttinger liquid description of charge and spin modes, as described above in Eq.~(\ref{eq:HchargespinLuttinger}). The Luttinger parameters read
\begin{align}
K_\rho &=  \sqrt{\frac{\Gamma}{4 v^0_F + \Gamma - 4 v^1_F }} ,\;  K_\sigma = \sqrt{\frac{\Gamma - 2 v^1_F (1 - \chi)}{\Gamma - 2 v^1_F (1 + \chi)} }, \nonumber \\
u_\rho &= \frac{\Gamma}{K_\rho}, \; u_\sigma = \sqrt{(\Gamma - 2 v^1_F (1 - \chi))(\Gamma - 2 v^1_F (1 + \chi) )}.
\end{align}

Finally, the form of $H_{i,2}$ in the Luttinger description is fixed by its action of changing $\mathcal{I}_z$ in steps of $2$, as in Eq.~(\ref{eq:Hi2}). The parameter $g$ is given by
\begin{align}
g &= \int \frac{d q_x}{2 \pi} \left[ F_{AB} (\vs{q})\right]^2 \cdot \nonumber \\
& \; \;\; \; \; \; \; \; \; \; \; \;   \left[ V (\vs{q} + \vs{K}_A - \vs{K}_B ) - V (\vs{q} + \vs{K}_A + \vs{K}_B ) \right]  \big|_{q_y = 0}. 
\end{align}

This completes our estimation of the parameters of the Luttinger theories. 

%

\end{document}